\documentclass{revtex4}
\usepackage{amsfonts}
\usepackage{amssymb}
\usepackage{amsmath}

\input{tcilatex}

\begin{document}

\title{Hawking Radiation of Massive Vector Particles From Warped AdS$_{\text{%
3}}$ Black Hole}
\author{H. Gursel$^{\ast }$}
\author{I. Sakalli$^{\dag }$}
\affiliation{Department of Physics, Eastern Mediterranean University,}
\affiliation{G. Magusa, North Cyprus, Mersin-10, Turkey.}
\affiliation{$^{\ast }$huriye.gursel@cc.emu.edu.tr}
\affiliation{$^{\dagger }$izzet.sakalli@emu.edu.tr}

\begin{abstract}
Hawking radiation (HR) of massive vector particles from a rotating Warped
Anti-de Sitter black hole in 2+1 dimensions (WAdS$_{\text{3}}$BH) is studied
in detail. The quantum tunneling approach with the Hamilton-Jacobi method
(HJM) is applied in the Proca equation (PE), and we show that the radial
function yields the tunneling rate of the outgoing particles. Comparing the
result obtained with the Boltzmann factor, we satisfactorily reproduce the
Hawking temperature (HT) of the WAdS$_{\text{3}}$BH.
\end{abstract}

\maketitle

\section{Introduction}

General relativity (GR) predicts a mysterious object, which is a spacetime
whose attractive gravitational force is so intense that no matter, light, or
information of any kind can escape. This mysterious object consists of a
point of no return known as the event horizon and a singularity with no
volume. In classical approach, unless the particle can move backwards in
time, which yet does not seem to be possible, it will not be able to get
out. Having John Wheeler \cite{Graviton} named these mysterious objects as
`black holes (BHs)', physicists are forced to ask this question to
themselves; are BHs really black? So far, HR \cite{Hawking1,Hawking2} seems
to be only calculable theoretical application of GR and quantum mechanics to
a BH. It analyzes the evaporation of a BH, which lasts for an infinitely
long period of time in an empty space. In addition to the HR, a wide range
of studies in theoretical and experimental physics shows that BHs emit
strong gravitational waves that is being hoped to prove their existence \cite%
{Gwaves}.

In this study, we aim to work out the temperature of HR emitted from a WAdS$%
_{\text{3}}$BH \cite{Warped,Warped2} using PE \cite{Krug} followed by HJM 
\cite{Review}. PEs are used to explore the behavior of the wave function of
spin-1 vector particles, which describe gauge bosons of the Standard Model:
force particles (e.g. weak massive $W^{\pm }$ and $Z^{0}$ bosons). These
vector particles can have their spin being pointed in any direction.
Recently, HR of the massive spin-1 vector particles tunneling from various
BHs has attracted much attention \cite{Krug2,Chin0,Chin,Chin2,Chin3,Ali}.
When HJM with the Wentzel--Kramers--Brillouin (WKB) approximation is
employed in PE, a coefficient matrix is obtained. By setting the determinant
of this matrix to zero, a set of linear equations can be solved for the
radial function. Consequently, one can compute the tunneling rate of the
gauge bosons, and derive the HT of WAdS$_{\text{3}}$BH.

In this paper, Section 2 is devoted to the introduction of WAdS$_{\text{3}}$%
BH. In Sec. 3, we analyze PE for a massive gauge boson field in this
geometry. We also show the the derivation of HR of the vector particles
tunneling from the WAdS$_{\text{3}}$BH. In Sec. 4, we conclude our
investigations.

Throughout the paper, we use units wherein $c=G=k_{B}=1$.

\section{WAdS$_{\text{3}}$BH Spacetime}

Conformal field theory (CFT) \cite{CFT} plays a vital role on the derivation
of WAdS$_{\text{3}}$BHs due to the AdS/CFT correspondence \cite{AdS/CFT}.
Its metric is the result of the Arnowitt-Deser-Missner (ADM) formalism with
integer spin bosons \cite{ADM,ADM2}. ADM is a topological trick based on the
decomposition of spacetime into space and time \cite{CARLIPLECTURE}. During
this process, a shift does not only occur in time but also in spatial
coordinates which, in our case, arises as the lapse function $N(r)$\ and the
shift vector $N^{\phi }(r)$. Furthermore, the geometry of WAdS$_{\text{3}}$%
BH has a symmetry breaking of $SL(2,%
%TCIMACRO{\U{211d} }%
%BeginExpansion
\mathbb{R}
%EndExpansion
)_{L}$ x $SL(2,%
%TCIMACRO{\U{211d} }%
%BeginExpansion
\mathbb{R}
%EndExpansion
)_{R}$ into $SL(2,%
%TCIMACRO{\U{211d} }%
%BeginExpansion
\mathbb{R}
%EndExpansion
)$x$U(1)$ \cite{Warped}.

Einstein-Hilbert action, which describes topologically massive gravity, is
given by \cite{CARLIPLECTURE}

\begin{equation}
I=\frac{1}{16\pi }\int_{M}d^{3}x\sqrt{-g}\left( 
%TCIMACRO{\U{211d} }%
%BeginExpansion
\mathbb{R}
%EndExpansion
-2\Lambda \right) +I_{matter},  \label{1}
\end{equation}

where $\Lambda ,$ $I_{matter},$ and $%
%TCIMACRO{\U{211d} }%
%BeginExpansion
\mathbb{R}
%EndExpansion
$ represent cosmological constant, gravitational Chern-Simons action and
Ricci tensor trace, respectively. In Schwarzschild coordinates, the
line-element of WAdS$_{\text{3}}$BH takes the following form \cite{Warped}

\begin{equation}
ds^{2}=-N^{2}dt^{2}+R^{2}\left[ d\phi +N^{\phi }dt\right] ^{2}+\frac{\ell
^{2}dr^{2}}{4R^{2}N^{2}},  \label{2}
\end{equation}

where 
\begin{eqnarray}
R^{2} &=&\frac{r}{4}\left[ 3\left( \nu ^{2}-1\right) r+\left( \nu
^{2}+3\right) \left( r_{+}+r_{-}\right) -4\nu \sqrt{r_{+}r_{-}\left( \nu
^{2}+3\right) }\right] ,  \label{3} \\
N^{2} &=&\frac{\left( \nu ^{2}+3\right) \left( r-r_{+}\right) \left(
r-r_{-}\right) }{4R(r)^{2}},  \label{4} \\
N^{\phi } &=&\frac{2\nu r-\sqrt{r_{+}r_{-}\left( \nu ^{2}+3\right) }}{%
2R(r)^{2}}.  \label{5}
\end{eqnarray}

Wick rotation \cite{WICK} being applied to two-dimensional spacetime
predicts holomorphic and anti-holomorphic functions that are stated as left
and right moving central charges. The corresponding right and\ left moving
temperatures, respectively, are given by \cite{Warped}

\begin{equation}
T_{R}=\frac{\left( \nu ^{2}+3\right) }{8\pi \ell }\left( r_{+}-r_{-}\right) ,
\label{6}
\end{equation}

\begin{equation}
T_{L}=\frac{\left( \nu ^{2}+3\right) }{8\pi \ell }\left( r_{+}-r_{-}-\frac{%
\sqrt{\left( \nu ^{2}+3\right) r_{+}r_{-}}}{\nu }\right) .  \label{7}
\end{equation}

Moreover,

\begin{eqnarray}
T_{H} &=&\frac{\kappa }{2\pi }=\left( \frac{T_{R}+T_{L}}{T_{R}}\frac{4\pi
\nu \ell }{\nu ^{2}+3}\right) ^{-1},  \notag \\
&=&\frac{\nu ^{2}+3}{4\pi \ell }\frac{r_{+}-r_{-}}{\left( 2\nu r_{+}-\sqrt{%
r_{+}r_{-}\left( \nu ^{2}+3\right) }\right) },  \label{8}
\end{eqnarray}%
\begin{equation}
\Omega _{H}=\frac{2}{\left( 2\nu r_{+}-\sqrt{r_{+}r_{-}\left( \nu
^{2}+3\right) }\right) \ell },  \label{9_ang}
\end{equation}

where $T_{H},$ $\kappa $, and $\Omega _{H}$ denote the HT, surface gravity
and angular velocity of WAdS$_{\text{3}}$BH. For each $\nu $ value, we have
a different case to be considered; having $\nu <1$ suggests squashed AdS$_{%
\text{3}}$, whereas $\nu >1$ implies AdS$_{\text{3}}$ to be stretched. Both
cases are spacelike and if the chiral point $\nu $ $=1$ is met, we can no
longer consider warpage \cite{Warped}. These BHs have very interesting
applications in the literature (see for instance \cite{App1,App2,App3,App4}
and references therein).

The special linear group $SL(n,%
%TCIMACRO{\U{211d} }%
%BeginExpansion
\mathbb{R}
%EndExpansion
)$ includes $n\times n$ matrices having determinants that are equal to one.

\section{Quantum Tunneling of Vector Particles From WAdS$_{\text{3}}$BHs}

For a massive vector boson (spin-1) field, PE in a curved spacetime is given
by \cite{Krug} 
\begin{equation}
D_{\mu }\Psi ^{\beta \mu }+\frac{m^{2}}{\hbar ^{2}}\Psi ^{\beta }=0,
\label{10}
\end{equation}

in which

\begin{equation}
\Psi _{\beta \mu }=D_{\beta }\Psi _{\mu }-D_{\mu }\Psi _{\beta }=\partial
_{\beta }\Psi _{\mu }-\partial _{\mu }\Psi _{\beta },  \label{11}
\end{equation}

where $\Psi _{\beta },$ $D_{\mu },$ $m$ and $\hbar $ represent three
component spinor field, covariant derivatives, mass and reduced Planck
constant, respectively. Since $\beta =\{0,1,2\}$ and $x^{\beta }=\{t,r,\phi
\}$, PEs can be expressed as a set of triad equations:

\begin{equation}
m^{2}\ell ^{2}R^{2}\Xi +\hbar ^{2}\ell ^{2}\partial _{\phi }(\Psi _{t\phi
})+4\hbar ^{2}R^{2}N^{2}\partial _{r}\xi =0,  \label{12}
\end{equation}

\begin{equation}
m^{2}R^{2}N^{2}\Psi _{r}+\hbar ^{2}(N^{\phi })^{2}R^{2}\lambda +\xi \hbar
^{2}=0,  \label{13}
\end{equation}

\begin{equation}
m^{2}\ell ^{2}R^{2}\Xi +\hbar ^{2}\ell ^{2}\partial _{\phi }(\Psi _{t\phi
})+4\hbar ^{2}R^{2}N^{2}\delta =0,  \label{14}
\end{equation}

where

\begin{eqnarray}
\Xi &=&\Psi _{t}-N^{\phi }\Psi _{\phi },  \notag \\
\xi &=&R^{2}\frac{\partial \Psi _{tr}}{\partial t}+N^{2}\frac{\partial \Psi
_{r\phi }}{\partial \phi },  \notag \\
\lambda &=&\frac{\partial \Psi _{r\phi }}{\partial t}-\frac{\partial }{%
\partial \phi }\left( \Psi _{tr}+\Psi _{r\phi }\right) ,  \notag \\
\delta &=&\partial _{r}(R^{2}\Psi _{tr})+\partial _{r}(N^{\phi }R^{2}\Psi
_{r\phi }).  \label{15}
\end{eqnarray}

Let us assume an ans\"{a}tz of the form

\begin{equation}
\Psi _{\nu }=\alpha _{\nu }\exp \left[ \frac{i}{\hbar }S(t,r,\phi )\right] ,
\label{16}
\end{equation}%
where $\alpha _{\nu }=\{\alpha _{0},\alpha _{1},\alpha _{2}\}$ represents
some arbitrary constants, and the action $S$ is defined as

\begin{equation}
S(t,r,\phi )=S_{0}(t,r,\phi )+\hbar S_{1}(t,r,\phi )+\hbar
^{2}S_{2}(t,r,\phi )+...  \label{17}
\end{equation}

Furthermore, we may set

\begin{equation}
S_{0}(t,r,\phi )=-Et+W(r)+j\phi +%
%TCIMACRO{\U{2102} }%
%BeginExpansion
\mathbb{C}
%EndExpansion
,  \label{18n}
\end{equation}

where $E$ and $j$ denote the energy and angular momentum of the spin-1
particle, respectively, and $%
%TCIMACRO{\U{2102} }%
%BeginExpansion
\mathbb{C}
%EndExpansion
$ is a complex constant. After applying the WKB approximation, we obtain the
elements of the coefficient matrix $H$ for the Eqs. (12)-(14), to the lowest
order in $\hbar $, as follows

\begin{eqnarray}
H^{00} &=&-4R^{4}N^{2}W^{\prime 2}-\ell ^{2}\left( m^{2}R^{2}+j^{2}\right) ,
\notag \\
H^{01} &=&H^{10}=-4R^{4}N^{2}W^{\prime }E_{net}  \notag \\
H^{02} &=&H^{20}=-\ell ^{2}jE+m^{2}\ell ^{2}R^{2}N^{\phi
}+4R^{4}N^{2}N^{\phi }W^{\prime 2},  \notag \\
H^{11} &=&-4N^{2}R^{2}\left[ \left( -m^{2}N^{2}+E_{net}^{2}\right)
R^{2}-N^{2}j^{2}\right] ,  \notag \\
H^{12} &=&H^{21}=4W^{\prime }\left[ NE_{net}R^{2}-N^{2}j\right] N^{2}R^{2}, 
\notag \\
H^{22} &=&\left[ -4R^{4}N^{2}\left( N^{\phi }\right) ^{2}+4R^{2}N^{4}\right]
W^{\prime 2}-\ell ^{2}\left[ E^{2}+m^{2}R^{2}\left( N^{\phi }\right)
^{2}-m^{2}N^{2}\right] .  \label{19n}
\end{eqnarray}%
where $W^{\prime }=\frac{d}{dr}W(r)$, and

\begin{equation}
E_{net}=E+jN^{\phi }.  \label{20n}
\end{equation}

Having non-trivial solution for the radial function $W_{\pm }(r)$ is
conditional on $\det (H)=0:$

\begin{equation}
\det (H)=-4R^{2}N^{2}m^{2}-4R^{4}N^{4}W^{\prime 2}+\ell ^{2}\left[ \left(
-m^{2}N^{2}+E_{net}^{2}\right) R^{2}-N^{2}j^{2}\right] ^{2}=0,  \label{21}
\end{equation}

and it is worthwhile to mention that $W_{+}$ stands for the radial function
of outgoing (emitted) particles, and conversely $W_{-}$ corresponds to the
ingoing (absorbed) solution. Therefore, the solution for the radial function
is

\begin{equation}
W_{\pm }(r)=\pm \int \frac{\ell }{2RN^{2}}\sqrt{E_{net}^{2}-N^{2}\left(
m^{2}+\frac{j^{2}}{R^{2}}\right) }dr.  \label{22}
\end{equation}

In the vicinity of the event horizon, $N$ approaches to zero, hence the
second term inside the square root vanishes, and the integral becomes

\begin{eqnarray}
W_{\pm }(r) &\approx &\pm \frac{\ell }{2}\int \frac{E_{net}}{RN^{2}}dr, 
\notag \\
&=&\pm \int \frac{2\ell E_{net}}{\left( \nu ^{2}+3\right) \left(
r-r_{+}\right) \left( r-r_{-}\right) }\sqrt{\frac{r}{4}\left[ 3\left( \nu
^{2}-1\right) r+\Upsilon \right] }dr,  \label{23}
\end{eqnarray}

where

\begin{equation}
\Upsilon =\left( \nu ^{2}+3\right) (r_{+}+r_{-})-4\nu \sqrt{r_{+}r_{-}\left(
\nu ^{2}+3\right) }.  \label{24}
\end{equation}

As it can be seen from Eq. (23), the integrand has a simple pole at the
event horizon, $r_{+}$. Using the Feynman's prescription \cite{FP}, we obtain

\begin{equation}
W_{+}(r)\cong \frac{i\pi \ell E_{net}\sqrt{3r_{+}^{2}\left( \nu
^{2}-1\right) r+\Upsilon r_{+}}}{\left( \nu ^{2}+3\right) (r_{+}-r_{-})}.
\label{25}
\end{equation}

Therefore, one can deduce that imaginary parts of the action can only come
about due to the pole at the horizon or from the imaginary part of $%
%TCIMACRO{\U{2102} }%
%BeginExpansion
\mathbb{C}
%EndExpansion
$. The probabilities of the particles crossing the event horizon each way
are given by \cite{Review}

\begin{equation}
\Gamma _{emission}=\exp (-\frac{2}{\hbar }\func{Im}S)=\exp [-\frac{2}{\hbar }%
\left( \func{Im}W_{+}+\func{Im}%
%TCIMACRO{\U{2102} }%
%BeginExpansion
\mathbb{C}
%EndExpansion
\right) ],  \label{26n}
\end{equation}

and

\begin{equation}
\Gamma _{absorption}=\exp (-\frac{2}{\hbar }\func{Im}S)=\exp [-\frac{2}{%
\hbar }\left( \func{Im}W_{-}+\func{Im}%
%TCIMACRO{\U{2102} }%
%BeginExpansion
\mathbb{C}
%EndExpansion
\right) ].  \label{27}
\end{equation}

According to the semi-classical approach, there exists a 100\% chance for
the ingoing spin-1 particles to enter the BH, i.e., $\Gamma _{absorption}=1$%
. This implies that $\func{Im}%
%TCIMACRO{\U{2102} }%
%BeginExpansion
\mathbb{C}
%EndExpansion
=-\func{Im}W_{-}.$ Since $W_{-}=-W_{+},$ whence we can read the probability
of the outgoing (tunneling) particles as

\begin{equation}
\Gamma _{emission}=\exp \left( -\frac{4}{\hbar }\func{Im}W_{+}\right) .
\label{28}
\end{equation}

Thus, the tunneling rate becomes

\begin{equation}
\Gamma _{rate}=\frac{\Gamma _{emission}}{\Gamma _{absorption}}=\exp \left( -%
\frac{4}{\hbar }\func{Im}W_{+}\right) ,  \label{29}
\end{equation}

which is also equivalent to the Boltzmann factor: $\exp (-E_{net}/T)$ \cite%
{Review,Chin,Ali2}. From this point on, we can compute the surface
temperature of the event horizon of WAdS$_{\text{3}}$BH as

\begin{equation}
T_{s}=\frac{\hbar }{2\pi }\frac{\left( \nu ^{2}+3\right) (r_{+}-r_{-})}{2l%
\sqrt{3r_{+}^{2}\left( \nu ^{2}-1\right) +\Upsilon r_{+}}}.  \label{30}
\end{equation}

Henceforth, we set $\hbar =1$, and Eq. (30) can be rewritten as

\begin{eqnarray}
T_{H} &=&\frac{\left( \nu ^{2}+3\right) (r_{+}-r_{-})}{4\pi l\sqrt{4\nu
^{2}r_{+}^{2}+\nu ^{2}r_{+}r_{-}+3r_{+}r_{-}-4\nu r_{+}\sqrt{%
r_{+}r_{-}\left( \nu ^{2}+3\right) }}},  \notag \\
&=&\frac{\left( \nu ^{2}+3\right) (r_{+}-r_{-})}{4\pi l\left( 2\nu r_{+}-%
\sqrt{r_{+}r_{-}\left( \nu ^{2}+3\right) }\right) }.  \label{31}
\end{eqnarray}

Equation (31) is fully agree with Eq. (8), which is the HT of WAdS$_{\text{3}%
}$BH.

\section{Conclusion}

The present study has constituted a valid process of evaluating HR by
promoting the application of PEs together with HJM. As an application, we
have shown how the quantum tunneling of vector particles from WAdS$_{\text{3}%
}$BH results in the standard HT, which is the combination of holomorphic and
anti-holomorphic functions; $T_{R}$ and $T_{L}$. We plan to extend our work
to higher dimensional rotating BHs. This is going to be our next work in the
near future.\


\begin{thebibliography}{99}
\bibitem{Graviton} C.W. Misner, K.S. Thorne, and J.A. Wheeler, \textit{%
Gravitation} (W. H. Freeman, San Francisco, 1973).

\bibitem{Hawking1} S.W.\ Hawking, Nature (London) \textbf{248,} 30 (1974).

\bibitem{Hawking2} S.W. Hawking, Commun. Math. Phys. \textbf{43,} 199
(1975); Erratum-ibid, \textbf{46}, 206 (1976).

\bibitem{Gwaves} M. Maggiore, \textit{The Detection of Gravitational Waves} 
\textit{Volume 1: Theory and Experiments} (Oxford University Press, Oxford,
2007).

\bibitem{Warped} D. Anninos, W. Li, M. Padi, W. Song, and A. Strominger,
JHEP \textbf{03,} 130 (2009).

\bibitem{Warped2} G. Cl\'{e}ment, Class. Quant. Grav. \textbf{26,} 105015
(2009).

\bibitem{Krug} S. I. Kruglov, Int. J. Mod. Phys. A \textbf{29,} 1450118
(2014).

\bibitem{Review} L. Vanzo, G. Acquaviva, and R. Di Criscienzo, Class.
Quantum Grav. \textbf{28,} 183001 (2011).

\bibitem{Krug2} S. I. Kruglov, Mod. Phys. Lett. A \textbf{29,} 1450203
(2014).

\bibitem{Chin0} \textbf{G. R. Chen and Y. C. Huang, Int. J. Mod. Phys. A 30,
1550083 (2015).}

\bibitem{Chin} G. R. Chen, S. Zhou, and Y. C. Huang, Int. J. Mod. Phys. D 
\textbf{24,} 1550005 (2015).

\bibitem{Chin2} G. R. Chen, S. Zhou, and Y. C. Huang, Astrophys. Space Sci. 
\textbf{357,} 51 (2015).

\bibitem{Chin3} \textbf{X. Q. Li and G. R. Chen, "Massive vector particles
tunneling from Kerr and Kerr-Newman black holes", arXiv:1507.03472.}

\bibitem{Ali} I. Sakalli and A. Ovgun, Eur. Phys. J. Plus \textbf{130}, 110
(2015).

\bibitem{CFT} J. Maldacena and A. Strominger, JHEP \textbf{9812}, 005 (1998).

\bibitem{AdS/CFT} E. Papantonopoulos (ed.), \textit{From Gravity to Thermal
Gauge Theories: The AdS/CFT Correspondence, Lecture Notes in Physics}
(Springer-Verlag, Berlin, Heidelberg, 2011).

\bibitem{ADM} R. Arnowitt, S. Deser, and C. Misner, Phys. Rev. \textbf{116,}
1322 (1959).

\bibitem{ADM2} S. Carlip, Living Rev. Rel. \textbf{8},1 (2005).

\bibitem{CARLIPLECTURE} S. Carlip, J. Korean Phys. Soc. \textbf{28}, S447
(1995).

\bibitem{WICK} D. Anninos, T. Anous, D. Z. Freedman, and G. Konstantinidis,
" \textit{Late-time Structure of the Bunch-Davies De Sitter Wavefunction}",
arXiv:1406.5490.

\bibitem{App1} M. Guica, T. Hartman, W. Song, and A. Strominger, Phys. Rev.
D \textbf{80,} 124008 (2009).

\bibitem{App2} D. Anninos, J. Samani, and E. Shaghoulian, JHEP \textbf{1402,}
118 (2014).

\bibitem{App3} K. Ait Moussa, G. Cl\'{e}ment, and C. Leygnac, Class. Quant.
Grav. \textbf{20,} L277 (2003).

\bibitem{App4} H. R. C. Ferreira and J. Louko, Phys. Rev. D \textbf{9,}
024038 (2015).

\bibitem{FP} N. N. Bogoliubov and D. V. Shirkov, \textit{Introduction to the
Theory of Quantized Fields} (John Wiley and Sons Ltd., 1980).

\bibitem{Ali2} I. Sakalli and A. Ovgun, "\textit{Hawking Radiation of Spin-1
Particles From Three Dimensional Rotating Hairy Black Hole}",
arXiv:1503.01316 (J. Exp. Theor. Phys+. \textbf{121}, No. 3 (2015); article
in press).
\end{thebibliography}
\end{document}